\documentclass[sigconf]{acmart}
\copyrightyear{2024}
\acmYear{2024}
\setcopyright{acmlicensed}\acmConference[ISWC '24]{Proceedings of the 2024 ACM International Symposium on Wearable Computers}{October 5--9, 2024}{Melbourne, VIC, Australia}
\acmBooktitle{Proceedings of the 2024 ACM International Symposium on Wearable Computers (ISWC '24), October 5--9, 2024, Melbourne, VIC, Australia}
\acmDOI{10.1145/3675095.3676620}
\acmISBN{979-8-4007-1059-9/24/10}

\usepackage{enumitem}
\setlist[itemize]{
    label=\textbf{\textbullet}, 
    leftmargin=*,
    align=left
}
\usepackage{booktabs}
\usepackage{graphicx}
\usepackage{caption}
\usepackage{subcaption}
\usepackage{multirow}
\usepackage{lipsum}  
\usepackage{caption}  
\usepackage{subcaption}  
\AtBeginDocument{%
  \providecommand\BibTeX{{%
    \normalfont B\kern-0.5em{\scshape i\kern-0.25em b}\kern-0.8em\TeX}}}

\begin{document}

\title{Understanding Physiological Responses of Students Over Different Courses}

\author{Soundariya Ananthan}
\orcid{0009-0002-4978-3360}
\affiliation{School of Computer Science and Engineering,
  \institution{University of New South Wales (UNSW)}
  \city{Sydney}
  \country{Australia}
}
\email{s.ananthan@unsw.edu.au}
\email{a.soundariya@gmail.com}

\author{Nan Gao}
\orcid{0000-0002-9694-2689}
\affiliation{Department of Computer Science and Technology,
  \institution{Tsinghua University}
  \city{Beijing}
  \country{China}
}
\affiliation{School of Computer Science and Engineering,
  \institution{University of New South Wales (UNSW)}
  \city{Sydney}
  \country{Australia}
}
\orcid{0000-0002-9694-2689}
\email{nangao@tsinghua.edu.cn}

\author{Flora D. Salim}
\orcid{0000-0002-1237-1664}
\affiliation{School of Computer Science and Engineering,
  \institution{University of New South Wales (UNSW)}
  \city{Sydney}
  \country{Australia}
}\email{flora.salim@unsw.edu.au}
\authornote{Corresponding authors}

\renewcommand{\shortauthors}{Ananthan, et al.}

\begin{abstract}
  Student engagement plays a vital role in academic success with high engagement often linked to positive educational outcomes. Traditionally, student engagement is measured through self-reports, which are both labour-intensive and not real-time. An emerging alternative is monitoring physiological signals such as Electrodermal Activity (EDA) and Inter-Beat Interval (IBI), which reflect students' emotional and cognitive states.  
  In this research, we analyzed these signals from 23 students wearing Empatica E4 devices in real-world scenarios. Diverging from previous studies focused on lab settings or specific subjects, we examined physiological synchrony at the intra-student level across various courses. We also assessed how different courses influence physiological responses and identified consistent temporal patterns. 
  Our findings show unique physiological response patterns among students, enhancing our understanding of student engagement dynamics. This opens up possibilities for tailoring educational strategies based on unobtrusive sensing data to optimize learning outcomes.

\end{abstract}

\ccsdesc[500]{Human-centered computing~Ubiquitous and mobile computing}
\ccsdesc[500]{Applied computing~Education}
\keywords{Engagement, Physiological Synchrony, Electrodermal Activity, Inter-beat Interval, Wearables, Courses}


\maketitle

\section{Introduction}
Engaging students in the learning process enhances their attention and focus, fostering a shift towards more critical thinking. Student engagement in the classroom is a better predictor of academic performance and overall learning outcomes \cite{Carini2006StudentEA}. Apart from academics, student engagement is also associated with their social and emotional outcome. Additionally, Christenson et al. \cite{Christenson2012HandbookOR} found that positive peer relationships influence student engagement in both academic and extracurricular activities. Numerous studies have explored the analysis of student engagement and behavioural changes using physiological signals, primarily focusing on understanding student behaviour and engagement within the classroom setting \cite{nan,Akhuseyinoglu,Lascio2018UnobtrusiveAO, Reading}. In the work of Jack et al. \cite{PE_ngage}, variation in students cognitive and affective processes has been detected with and without Physical Education (PE) classes during the school day. The findings from this research show the influence of a particular course on engagement.

Several studies \cite{Lo2021StudentEI, GERSAK2020581,Alkabbany} highlight the importance of detecting student engagement in mathematics, a crucial component of STEM (Science, Technology, Engineering, and Mathematics). They argue that promoting engagement in mathematics is vital for retaining students in STEM fields,  thereby contributing to the sustainable development of society. However, while existing research often concentrates on student behaviour and engagement within specific courses, there is a notable research gap in comparative analyses across different courses. Addressing this gap, our study aims to explore variations in student engagement by examining physiological signals across various courses. Specifically, we measure \textit{physiological synchrony} (PS) \cite{Gashi2019UsingUW}  among students to identify patterns of engagement that may differ from one course to another.

Physiological synchrony refers to the relationship between individuals physiological responses often occurring in social or interpersonal contexts. Traditionally, studies on the Autonomic Nervous System (ANS), focuses on synchrony within individuals to explore the temporal changes within them. Later, it has been identified in research the that autonomic process of one individual can affect or synchronize with another individual during social interactions \cite{Palumbo2017InterpersonalAP}. In our work, physiological synchrony is calculated at an intra-student level for diverse courses.  Intra-student involves finding the synchrony of a single student across different times of the same course. Examining synchrony within the same student helps to understand how an individuals physiological responses vary across different contexts, activities, or subjects. Furthermore, our investigation extends to the analysis of Electrodermal Activity (EDA) and Heart Rate (HR) features collected through wearables, revealing a significant relationship.
 
In this paper, the research questions are outlined as follows: \textit{1. Does the physiological synchrony of students vary across different courses? 2. Can different course characteristics influence students physiological responses? Are there consistent patterns in the temporal relationship across various courses?} To achieve this, we analyzed the EDA and IBI signals obtained from 23 students using the Empatica E4 device. To summarize, the details of our contribution are outlined below:
 \begin{itemize}[label=\textbf{\textbullet}]
    \item 	We examined the physiological synchrony of the students in different courses at intra-student level. To the best of our knowledge, we are the first to analyze physiological synchrony across different courses. Studying this phenomenon could help identify individual engagement and overall classroom environment.
    \item 	We extracted various EDA and HR features and formulated the relationship between them to understand the variation in physiological signals among different courses. Our results demonstrate variations in various features of physiological signals, suggesting that students exhibit different physiological responses at different times.
\end{itemize}

\section{Related works} \label{sec: Related work}

\subsection{Student Engagement}

Engagement is a malleable construct that can be influenced by personal factors, context, or stimuli and they change over time \cite{l2classroom}.  In \cite{Aro}, the author emphasizes the importance of investigating individual differences in engagement and how engagement varies among individuals. Zhou et al. \cite{zhou} conducted a study on student interaction in online learning based on an after-video quiz. They stated that students individual behaviour impacts both video-control behaviours and quiz scores.  Zhou et al.\cite{l2classroom} developed a longitudinal design to capture student engagement and disengagement within specific contexts over time. The study revealed a negative correlation between student engagement and disengagement levels when comparing the initial and final days of the semester indicating that student engagement changes over time. 

Cognitive ability and personality traits play a crucial role in determining individual performance and satisfaction. In line with the findings of Lall\'{e} et al. \cite{Lall}, it is evident that these traits have a direct impact on individuals performance, whether it be in real-world scenarios or artificial tasks. In the study conducted by Gao et al. \cite{classroom}, it was found that students who sit near each other have similar levels of learning engagement. Furthermore, the researchers observed a statistical correlation between individual and group seating arrangements in the classroom with both perceived and physiologically measured student engagement.

\subsection{Physiological Synchrony: A Comprehensive Analysis}

Physiological synchrony is observed in various scenarios and social contexts. In particular, they are highly found in parent-child relationships \cite{Feldman2007ParentinfantSA}, student-student interactions \cite{classroom}, teacher-student collaboration \cite{Lascio2018UnobtrusiveAO}, therapist-client sessions \cite{Ramseyer2011NonverbalSI}, and among couples \cite{spouse_interaction}. In the study by \cite{Lascio2018UnobtrusiveAO}, Lascio et al. captured the synchrony between students and the teacher which is used as one of the features to distinguish between engaged and non-engaged students. They used Dynamic Time Warping (DTW), Pearson Correlation (PC), and Single Session Index to capture physiological synchrony between teachers and students. 

Gashi et al. \cite{Gashi2019UsingUW} focused on analyzing the experience of presenters and the audience during the presentation by exploring their physiological synchrony. Researchers suggested that using the Dynamic Time Wrapping algorithm showed a positive correlation between physiological synchrony and participants self-reported scores. 
In the paper \cite{classroom},  Gao et al. reveal that there is a link between classroom seating and physiological synchrony.  They observed that students who sat close together had higher synchrony scores than others. The study \cite{Malmberg2019Are} investigated physiological synchrony at individual and group levels during collaborative exams. Moreover, their findings reveal that the groups who had difficulties showed synchrony in their physiological response.

\subsection{Physiological Signals Variations across Time}

Physiological signals vary for each individual and change based on different situations or activities. When it comes to students, physiological signals can be influenced based on their interests, prior knowledge, and challenges. There are various research papers that focus on the impact of individual courses on students physiological responses. In \cite{Rodríguez}, Rodríguez et al. described the difference in sustained attention of students, observing higher activation of the Sympathetic Nervous System (SNS) in the morning group than in the evening group. They also suggest that a physiological response that promotes attention is activated within the first 15 to 20 minutes of a conventional class. The study by Phung et al. \cite{Phung2017TaskPA} observed variation in student engagement when they performed their preferred task and dispreferred class. 

Fogarty et al. \cite{PE_ngage} observed that autonomic arousal is induced by PE class and they lasted approximately two hours after PE. Additionally, they found the arousal to be higher during the beginning of the class which subsides later. The finding emphasizes the need to understand the variation of physiological signals during different timings. Similarly, the effect of prayer, meditation and yoga in physiological response has also been studied widely \cite{Doufesh2013AssessmentOH,Cowen2007HeartRI,Wenger2007StudiesOA,Trivedi,Bernardi,Stigsby1981ElectroencephalographicFD}. The study by Subhadarshini et al. \cite{Subhadarshini2017Analysis} focuses on the impact of national anthem in the ANS. The participants showed the dominance of parasympathetic activity over sympathetic activity which indicates the state of rest or relaxation. Galan et al. \cite{Galan} used EEG (electroencephalography) to measure the cognitive workload when solving a math problem. They analyzed this by making students solve both easy and difficult math problems.

In summary, our approach differs from other studies that primarily examine physiological synchrony within specific contexts or natural environments. Furthermore, we assess various EDA and HR features across different courses, enabling us to identify trends within these courses. Statistically significant correlations between EDA and HR features have been discovered.

\section{Data Collection} \label{sec: data collection}

We utilized the publicly available dataset \textit{En-Gage} \footnote{The dataset download link: \url{https://physionet.org/content/in-gauge-and-en-gage/1.0.0/}} \cite{gao2022understanding}, which is a cross-sectional study conducted in a school in Melbourne, Australia. This dataset was employed for investigating student engagement and behaviour.  The dataset comprises physiological information such as electrodermal activity (EDA), heart rate (HR), blood volume pulse (BVP), skin temperature (ST), and 3-axis acceleration (ACC) collected using Empatica E4 wristband. Additionally, it includes self-reported engagement, thermal comfort, seating location, and emotion data collected from 23 students in a K-12 classroom, as well as 6 teachers, over a four-week period. The participants were selected on a voluntary basis and signed a consent form. In the case of student participants, their guardians signed the consent form.

The students were organized into three form groups for English, Science, and Politics, with a total of 10 different courses available: Assembly, Chapel, English, Form, Health, Language, Maths, PE, Politics, and Science. For our analysis, we have utilized EDA, IBI signals, and class details. The class details encompass various elements, namely the class ID, room allocation, date, start and end times, class length, week number, weekday, class number, course title, and classroom arrangement. As a gesture of appreciation for their involvement, each participating student received a certificate of participation and four movie vouchers, one for each week of successful participation. It is important to note that participation in this research project was voluntary, and participants were made aware that they could withdraw from the project at any stage. The data collection process received approval from the Science, Engineering and Health College Human Ethics Advisory Network (SEH CHEAN) at RMIT University. Additionally, the principal of the school where the study took place granted their approval for the project.

\section{Feature Extraction} \label{sec: feature extraction}
\subsection{Data Preprocesssing}
Before computing the synchrony score, the EDA signal is cleaned using Neurokit, an open-source Python package for neurophysiological signal processing \cite{neurokit}. In this study, the EDA signal is filtered using a Butterworth filter with a cutoff frequency around 0.05 Hz and a band-pass filter with a lower cutoff frequency of 0.1 Hz and an upper cutoff frequency of 5 Hz. Subsequently, the signal is decomposed into two components phasic and tonic. To identify the tonic component, the EDA signal is smoothed using a low-pass filter, and the peak values are estimated. After estimating the tonic component, the phasic component is extracted by subtracting the estimated tonic component from the original signal. Furthermore, artifact removal and baseline correction are done to remove any unwanted noise or residue offset in the signal.

\subsection{Feature Extraction based on Physiological Synchrony} \label{sec:physiological synchrony}

 To capture the physiological synchrony, we employed the Fast Dynamic Time Wrapping Algorithm (FastDTW) \cite{FastDTWTA,Gashi2019UsingUW} method which is used for measuring the similarity between time series data. We chose the FastDTW method as it can be run on larger datasets and is computationally efficient. Moreover, FastDTW is considered to be better than traditional DTW algorithms due to its efficiency, scalability, noise robustness, and memory efficiency. We applied FastDTW to three types of EDA data: EDA, Cleaned EDA, and Phasic EDA. After careful consideration, we opted to utilize the Phasic EDA data for our detailed analysis as they tend to show physiological and emotional arousal \cite{Critchley2013}. The higher the synchrony score among the two signals, the lower the synchrony and vice versa because FastDTW emphasizes the distance between two signals \cite{Gashi2019UsingUW}. FastDTW has been calculated for all the 10 courses - \textit{Assembly, Chapel, English, Form, Health, Language, Maths, Politics, PE, and Science}. 

\subsection{Features of Electrodermal Activity (EDA)}

EDA signals reflect the activity of sympathetic nerve traffic which is closely linked to the change in the mental state of the individual \cite{Critchley2013}. In our study, we focused on analyzing the statistical and event-related features of EDA. To accomplish this, we computed the mean, standard deviation, and variance of Skin Conductance Response (SCR), including SCR peak amplitude mean. Shukla et al. \cite{shukla} identified SCR peak amplitude as the most significant feature in determining valence and arousal. We also examined additional SCR features such as changepoint and autocorrelation \cite{halem}. Changepoint is used to detect the number of changes in the statistical properties of the EDA signal which indicates events or transitions. On the other hand, autocorrelation measures the similarity between the signal and its time-lagged version revealing temporal dependencies of the signal.

\subsection{Features derived from Inter-Beat Interval (IBI)} \label{sec: hr features}

Heart Rate Variability (HRV) is primarily used to assess the function of ANS which consists of SNS and PNS. The SNS is responsible for the fight-or-flight response, while the PNS is responsible for the rest-and-digest mechanism \cite{Ishaque2021TrendsIH}. Time and frequency domain features are commonly used to evaluate the performance of the ANS. Time domain features indicate changes in HR activity resulting from PNS or SNS activity. In our study, we extracted these features using pyhrv \cite{Gomes2019}, an open-source Python toolbox. Time domain features were extracted from the R peak and we computed metrics such as the mean, standard deviation of HR, NNI (Normal-to-Normal Interval), SDNN (Standard deviation of NN intervals), and RMSSD (Root mean of squared NNI differences). For frequency domain features, the Welch analysis method is applied to extract the following features: very low-frequency (VLF) peak, VLF absolute power, low-frequency (LF) peak, LF absolute power, very high-frequency (VF) peak, VF absolute power, LF/HF ratio, and total power. HF component indicates increased PNS activity which is commonly associated with periods of rest, recovery, and a state of relaxation. On the other hand, the LF component is linked to both SNS and PNS activities, reflecting stress or moderate physical activity. Lastly, the VLF component provides insights into variations in the overall ANS activity over extended periods.

\section{Analysis} \label{sec: analysis}

We initiated our investigation by examining the levels of physiological synchrony in various academic courses. Our null hypothesis is \textit{"There is no statistically significant difference in the level of physiological synchrony among students across different types of academic courses"}. To assess the validity of this hypothesis, we employed the analysis of variance (ANOVA) method on calculated physiological synchrony.  The results showed a statistical value of 3.711 and a P-value of 3.163e-07, which is less than 0.05. 
Since the ANOVA test is significant, we applied \textit{Tukey's Honestly Significant Difference} (HSD) to identify which specific groups are different from each other. We set the \textit{Family-wise Error Rate} (FWER) to 0.05, representing the probability of making a false positive. In other words, the HSD test attempts to ensure that, with 95\% confidence, none of the reported statistically significant differences are simply due to chance.

\begin{table}[h!]
\caption{Tukey's HSD Test Results for Statistically Significant Course Pairs}
\label{tab: statistical measures}
\resizebox{\columnwidth}{!}{%
\begin{tabular}{@{}lllllll@{}}
\toprule
\textbf{Course 1} & \textbf{Course 2} & \textbf{Mean diff} & \textbf{P-adj} & \textbf{Lower} & \textbf{Upper} & \textbf{Reject} \\ \midrule
Assembly & PE & 160.39 & 0.003 & 33.92 & 286.8 & True \\
English & PE & 141.10 & 0.016 & 14.64 & 267.6 & True \\
Language & PE & 144.28 & 0.012 & 17.82 & 270.7 & True \\
Maths & PE & 128.14 & 0.044 & 1.69 & 254.6 & True \\
PE & Politics & -128.31 & 0.044 & -254.8 & -1.85 & True \\
PE & Science & -131.13 & 0.035 & -257.6 & -4.68 & True \\ \bottomrule
\end{tabular}}
\end{table}

In total, we had 45 pairs of courses, out of which 6 have significant differences. Table \ref{tab: statistical measures} highlights six statistically significant differences between course pairs, marked by "True" in the \textit{Reject} column. These differences have adjusted p-values less than 0.05, indicating that they stand out as statistically significant after HSD adjustment for multiple comparisons. The \textit{Meandiff} column quantifies the difference in mean scores between the compared courses. The \textit{P-adj} column displays the adjusted p-values after HSD. These values, all below 0.05, confirm that the observed differences are not likely due to chance and are statistically significant. The \textit{Lower} and \textit{Upper} columns represent the 95\% confidence interval for the mean difference.  In summary, the 6 pairs with significant differences suggest that there are statistically significant variations between the corresponding groups. It can also be seen that all the six pairs involved the course PE.

\section{Results and Discussions} \label{sec: results}
\subsection{Physiological Synchrony of Students vary across Different Courses} \label{sec: RQ1}

\begin{figure}[h!]
  \centering
    \centering
    \includegraphics[width=0.9\columnwidth]{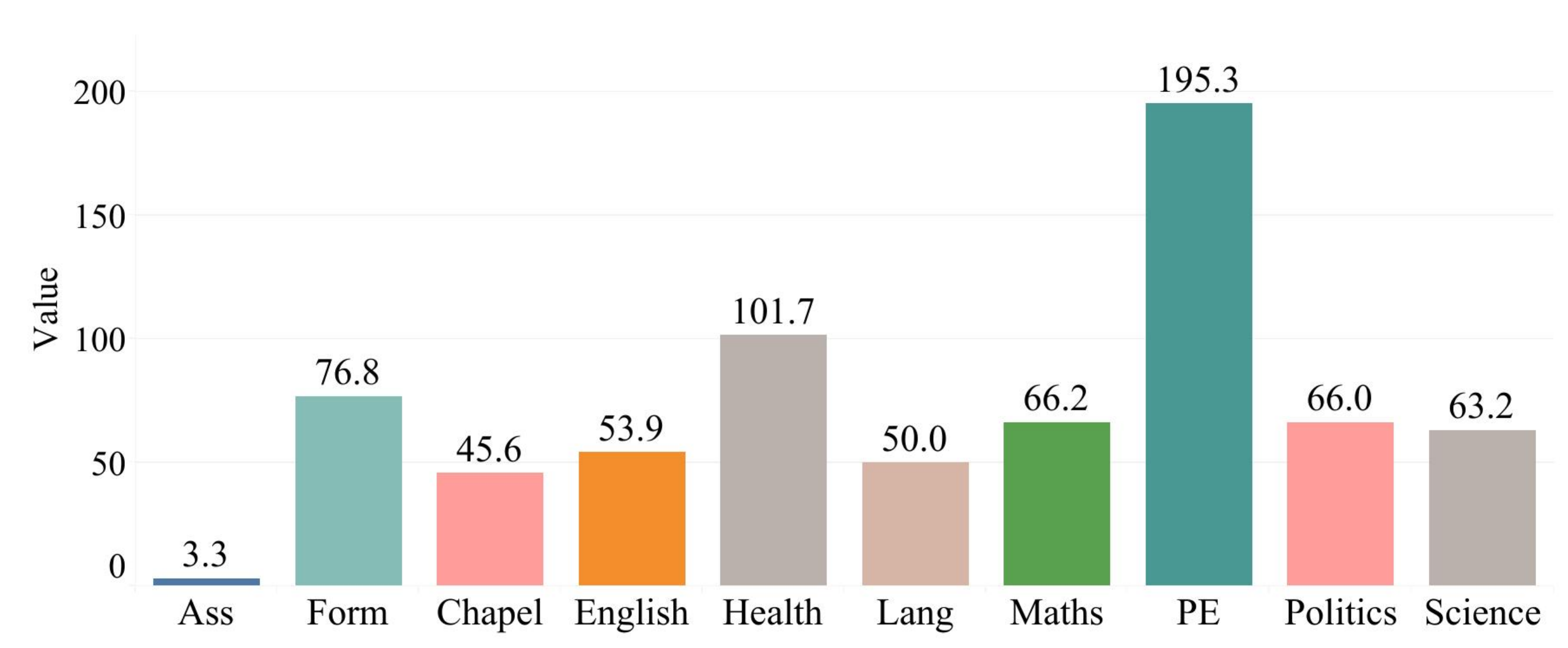}
    \caption{Physiological Synchrony of Students}
    \label{fig:Ps for intra-student}
    \Description{Figure shows the calculated synchrony score for all the ten courses. Apart from Assembly and PE courses, all the other courses have synchrony scores in a similar range which may be due to the subjective nature of the courses.}
\end{figure}

As mentioned earlier in Section \ref{sec:physiological synchrony}, FastDTW has been utilized to calculate physiological synchronies for all the courses.  Based on our data, the course Assembly has the lowest physiological synchrony value with score 3.3. On the other hand, the course PE showcases the highest synchrony score of 195.3, surpassing all other courses. The difference in physiological synchrony among all the courses can be observed in Figure \ref{fig:Ps for intra-student}. 
Interestingly, apart from Assembly and PE, all other courses show somewhat similar scores. 

The course Assembly involves activities such as gathering students and staff, singing the national anthem, and making announcements, while PE incorporates physical activities and exercises.
The observed physiological synchrony aligns with the nature of each course. The low scores for Assembly may be attributed to consistent and predictable physiological reactions, possibly due to the uniform nature of Assembly activities. On the other hand, the high FastDTW score for PE can be attributed to the diverse physical activities offered, which vary based on individual fitness levels.

\begin{figure}[h!]
    \centering
    \includegraphics[width=\linewidth]{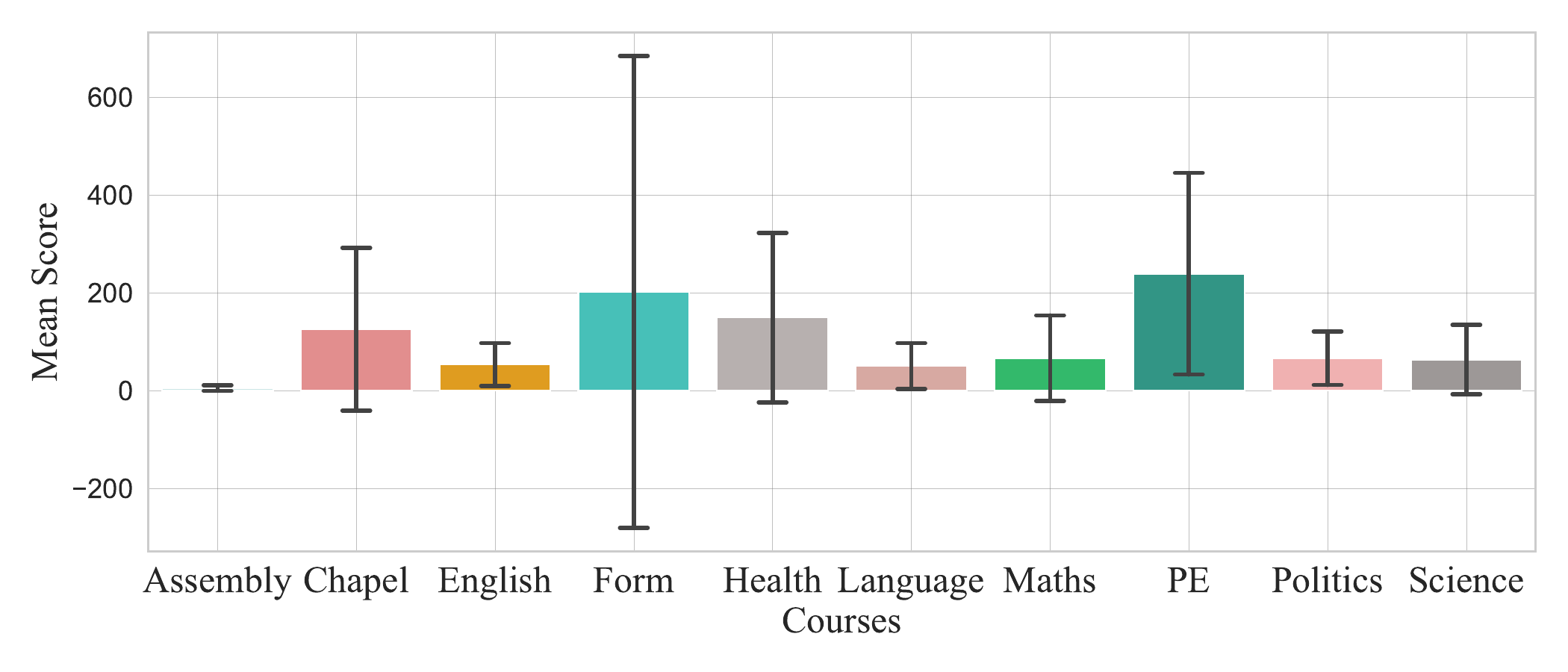}
    \caption{Error bars for all the Students}
    \label{fig:error for intra-student}
    \Description{Figure displays the predicted error bars for all the ten courses. It can be seen that Assembly has very less prediction errors. Even though PE marks high prediction error both PE and Form have noticeably high errors.}
\end{figure}

 Figure \ref{fig:error for intra-student} displays a box plot illustrating the absolute prediction error for each course. By analyzing prediction errors for each course, we can assess how individual courses influence the overall variability of PS. Course Assembly has lower prediction errors due to the predictable patterns of physiological synchrony enabling more accurate predictions. On the other hand, course PE has higher prediction errors due to the variations in physiological synchrony among individuals based on their fitness levels and engagement in different physical activities.

 \begin{figure}[h!]
 \centering
 \includegraphics[width=0.7\columnwidth]{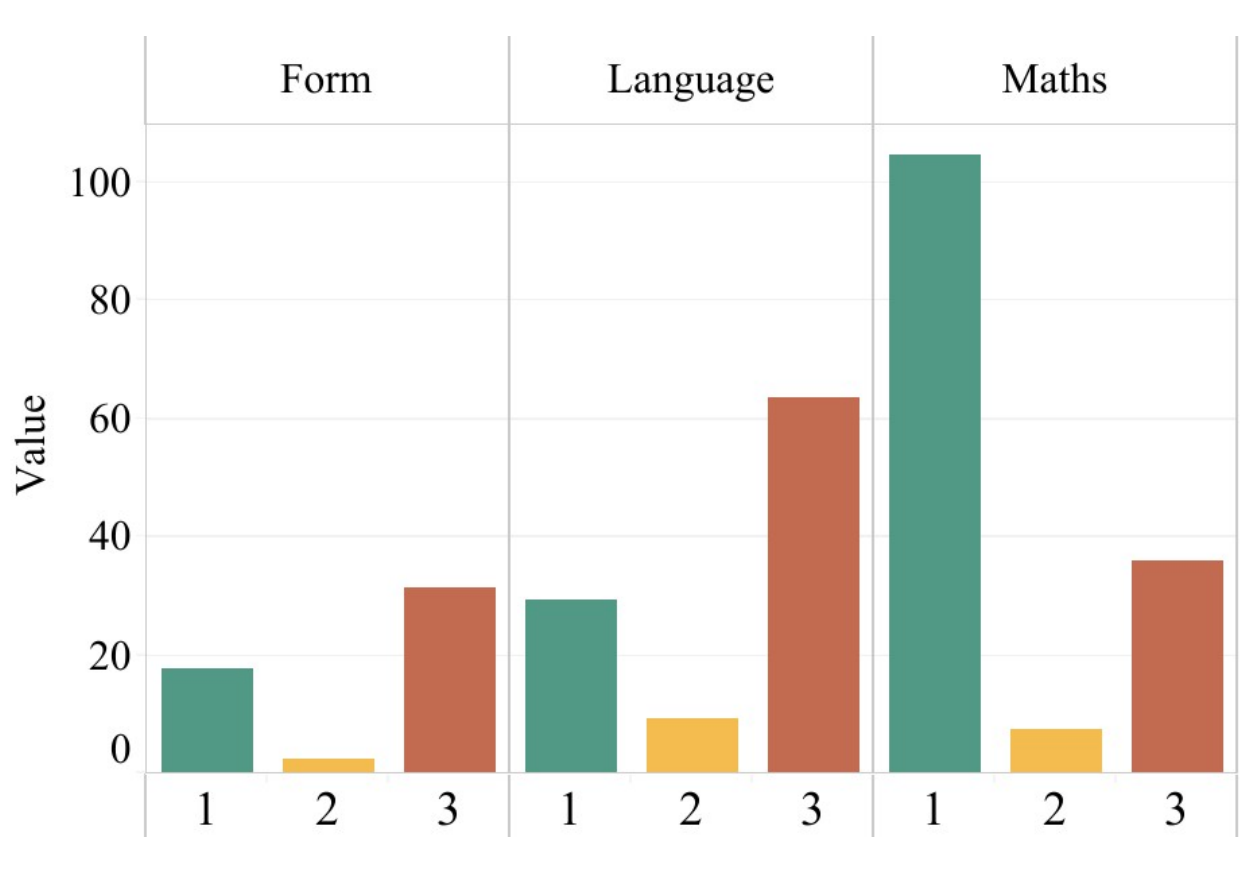}
 \caption{Student-wise PS scores}
 \label{students PS score}
 \Description{It can be seen that student 2 has low scores in all three courses. Student 1 has a lower synchrony score for Form and Language compared to Student 3 but shows a higher synchrony score for the course Maths. This demonstrates the variation in physiological signals among individuals.}
 \end{figure}

As explained in Section \ref{sec: Related work}, each individual has a unique physiological pattern that is influenced by their cognitive and personal factors. This phenomenon is evident in our study, as shown in Figure \ref{students PS score}. Figure \ref{students PS score} displays the synchrony scores of three different students (Student IDs 1, 2, and 3) for the courses of Form, Language, and Maths. Each student exhibits a unique pattern of synchrony scores, thus emphasizing the individualized nature of physiological responses within an educational setting. Recognizing and accommodating these individual differences can lead to more effective and targeted learning experiences. For example, collaborative learning environment can be created for students with synchronized physiological responses thereby enhancing their engagement.

\subsection{Relationship between Physiological Signals and the Courses} \label{sec: RQ2}

Now, we investigate whether different courses display different physiological signals. Our hypothesis here is that the course with high intensity will exhibit greater fluctuations in physiological responses than the course with stable or consistent workloads.

\subsubsection{Temporal Analysis of EDA Data}

We computed changepoint, autocorrelation, SCR peak and SCR peak amplitude mean measures individually for each course. The calculated changepoint and autocorrelation values from the EDA data show high values for PE  and low values for Assembly, indicating different patterns or transitions in the EDA data between the two courses. In [2], it is described that high arousal is associated with fewer peaks, low changepoint, and low autocorrelation values. On the other hand, positive energy (energy and enthusiasm) is associated with more peaks, low changepoint, and low autocorrelation values. 

\begin{figure}[H]
  \centering
  \includegraphics[width=0.9\columnwidth]{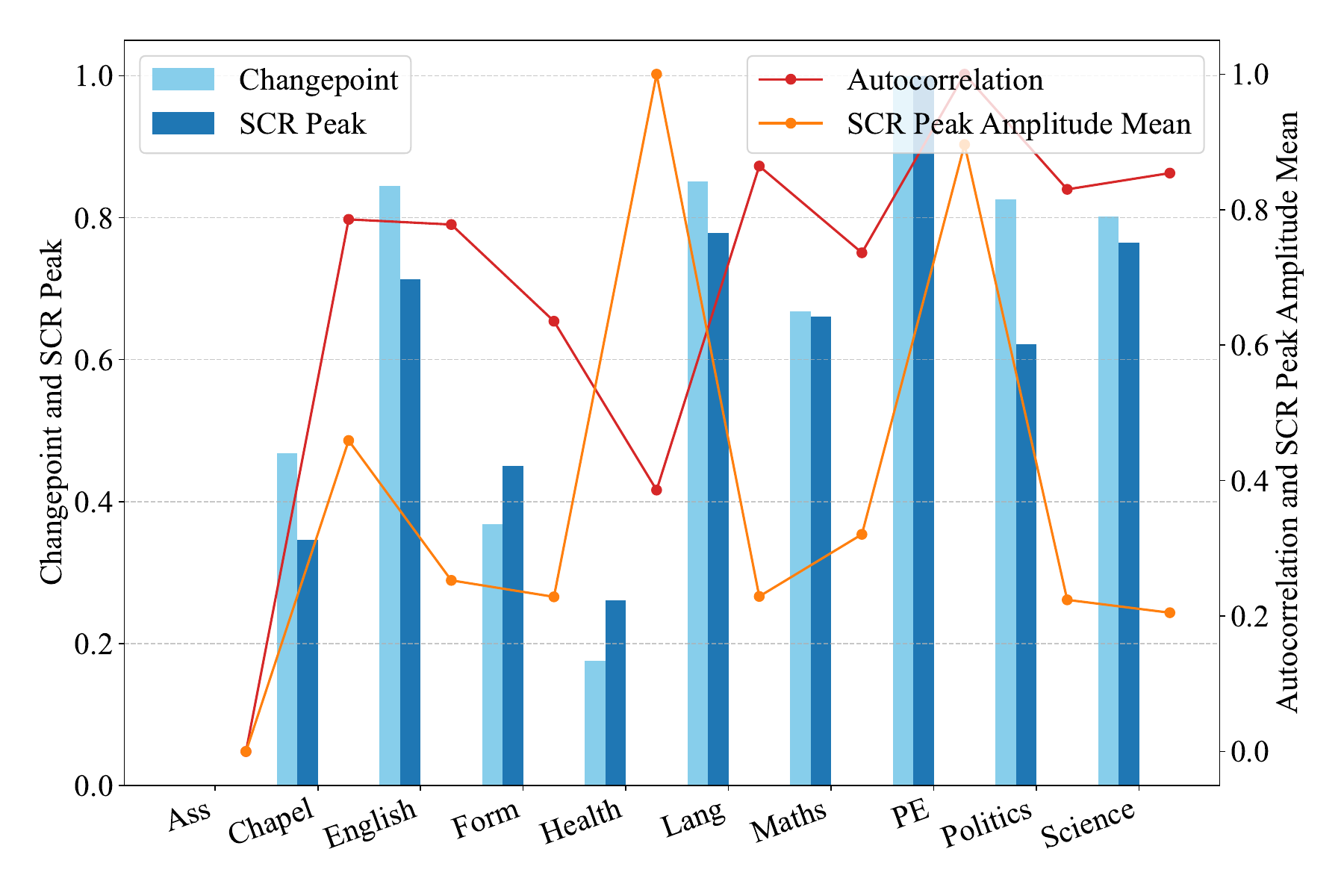}
  \caption{Changepoint and Autocorrelation Analysis}
  \label{fig:changepoint and autocorrelation}
  \Description{Course PE exhibits high changepoint, autocorrelation, and SCR peak and SCR peak amplitude mean indicating varying physiological responses. These responses are related to high energy or excitement during the activities. Conversely, low values of Assembly indicate a consistent pattern of physiological signal, signifying a relaxed form}
\end{figure}

Figure \ref{fig:changepoint and autocorrelation} displays the normalized changepoint, autocorrelation, SCR peak, and SCR peak amplitude values. It is evident that Assembly exhibits fewer peaks, low changepoint, and low autocorrelation values, indicating high arousal. This suggests that the EDA signal of Assembly is more temporally consistent or structured, while PE may be linked to changes in physical activity, resulting in more fluctuations in the EDA data.  

\subsubsection{Shift Point Identification in Physiological Signals}

During our investigation of students physiological responses in different classes, we discovered interesting heart rate patterns, specifically in Assembly and PE classes. In the Assembly class, an average heart rate of 109.4 beats per minute (bpm) is observed, which is remarkably high. Conversely, PE which typically involves physical activity, results in higher heart rates with an average HR of 101.3 bpm. The increased mean heart rate during Assembly may be due to factors such as stress, cognitive load, or other non-physical aspects associated with the course. This is analogous to the stress that students commonly experience during examinations, which has the potential to impact their heart rate. This approach significantly contributes to a more comprehensive understanding of the observed cardiovascular activity patterns. By drawing this parallel, it becomes apparent that heart rate dynamics are influenced by numerous factors beyond mere physical exertion, thereby underscoring the complex nature of this phenomenon. To further examine heart rate variability, we analyzed the Normal-to-Normal Interval (NNI) data. We observed that both the Assembly and PE classes had relatively shorter NNI intervals compared to the other courses, with figures of 615 for Assembly and 625 for PE as depicted in Table \ref{tab: HR and NNI}.

\begin{table}[h!]
\caption{Comparative Analysis of Mean Heart Rate and NNI Across Different Courses}
\label{tab: HR and NNI}
\begin{tabular}{@{}lll@{}}
\toprule
\textbf{Courses} & \textbf{Heart rate} & \textbf{NNI} \\ \midrule
\textit{Assembly} & 109.38 & 615.47 \\
\textit{Chapel} & 78.15 & 797.44 \\
\textit{English} & 80.15 & 778.99 \\
\textit{Form} & 81.36 & 770.69\\
\textit{Health} & 93.64 & 709.05 \\
\textit{Language} & 81.72 & 775.30 \\
\textit{Maths} & 79.28 & 784.78\\
\textit{PE} & 101.28 & 625.24 \\
\textit{Politics} & 78.77 & 790.82 \\
\textit{Science} & 80.69 & 776.14 \\ \bottomrule
\end{tabular}
\end{table}

The examination of NNI interval for the Assembly course in relation to changepoints introduced a captivating element to the investigation. The NNI interval for the Assembly course showed a higher value compared to the changepoint, while the relationship was reversed for the other courses.  
In the case of Assembly, NNI intervals were observed to be longer than changepoints, indicating a longer duration between consecutive normal heartbeats (NNI intervals) compared to the points at which significant changes occur (changepoints) in the EDA data. On the other hand, for the other courses, the pattern was the opposite, with NNI intervals being shorter than changepoints. This suggests that the time interval between heartbeats is shorter than the intervals between significant changes in the EDA data. This inconsistency between Assembly and the other courses may indicate variations in the relationship between physiological responses (HRV) and changes in physiological arousal (EDA changepoints) for these specific courses.

\subsubsection{Sympathetic and Parasympathetic Activation Across Academic Courses.}

Our investigation led to a significant finding of a positive correlation between the RMSSD and the peak amplitude of the SCR for the Assembly course with a correlation coefficient (r) of 0.54. This suggests that as RMSSD increases (indicating higher heart rate variability), SCR peak amplitudes also tend to increase. In other words, there is a relationship where changes in heart rate variability are associated with changes in SNS activity, as reflected by SCR. 

\begin{figure}[!h]
  \centering
  \includegraphics[width=0.8\columnwidth]{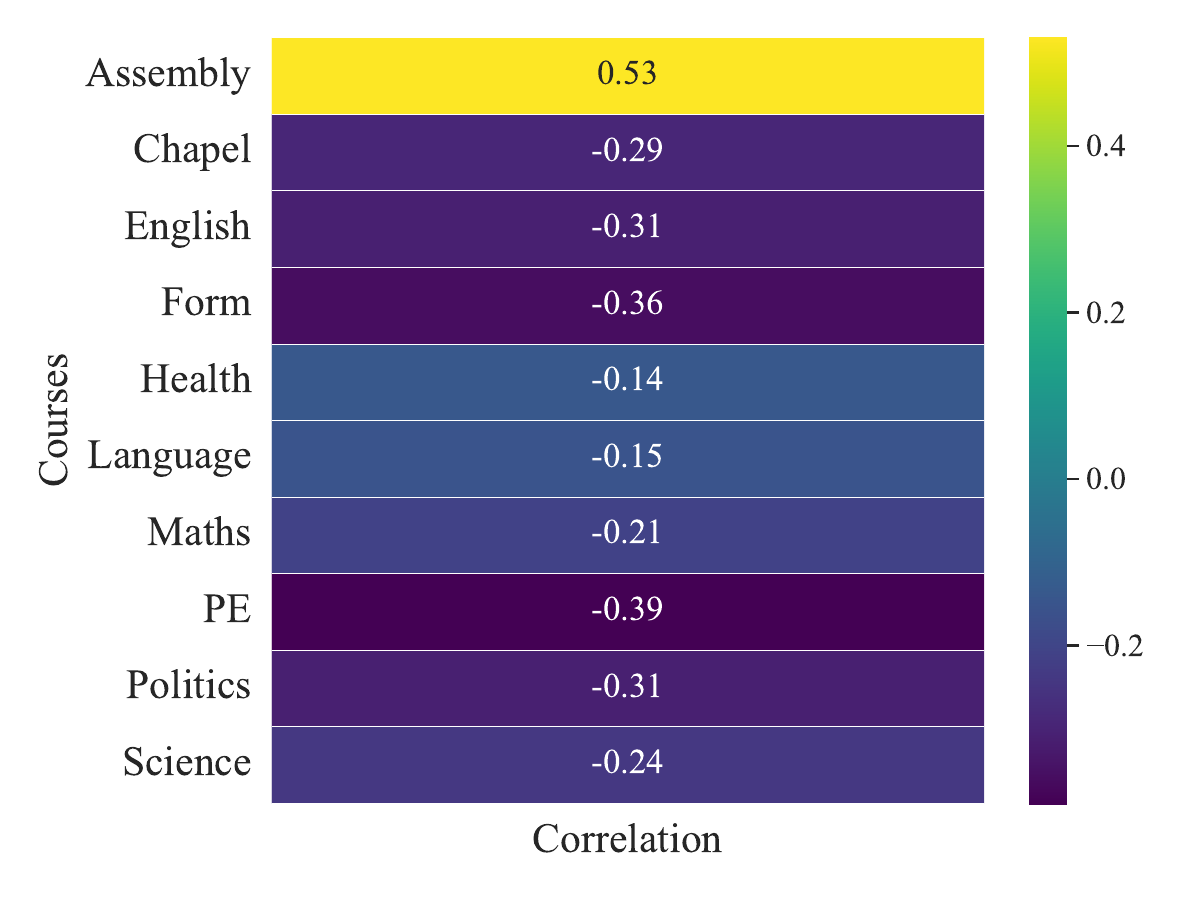}
  \caption{Time domain features}
  \label{fig:RMSSD and SCR}
  \Description{Correlation between RMSSD and SCR peak amplitude mean is positive only for the course Assembly. This correlation shows that there exists a relationship between PNS and SNS activity for the course Assembly which is not observed in any other courses.}
\end{figure}

The research by Yugar et al. \cite{Yugar} has shown that the activation of the SNS typically reduces heart rate variability. However, in the context of the Assembly course, the increase in both RMSSD and SCR peak amplitudes suggests the activation of the PNS. This aligns with a study \cite{Subhadarshini2017Analysis}  demonstrating that listening to the national anthem led to an increase in PNS activity and a decrease in SNS activity, which corresponds to our finding of increased HRV during the Assembly course. Figure \ref{fig:RMSSD and SCR} illustrates a correlation between the RMSSD and the peak amplitude of the SCR for all the courses. The positive relationship between RMSSD and SCR peak amplitudes challenges conventional expectations and emphasizes the need to interpret ANS dynamics in a context-specific manner.

Our study extends beyond the examination of time domain parameters to include a comprehensive analysis of the frequency domain parameters of HR features. As discussed in \ref{sec: hr features} activation of PNS is primarily identified by HF value and VLF indicates the SNS activity whereas LF is mostly used to indicate SNS activity but can specify PNS activity.  In our study, high activation of VLF and HF is found for the course Assembly indicating the presence of both PNS and SNS activity. Conversely, PE had the lowest HF value, indicating a lower level of PNS activity caused by physical activity. Our analysis shows that there is a variation in the physiological signals of students during different activities. Monitoring these patterns can be valuable in understanding an individual's physiological response to different stimuli and conditions.

\section{Implications and Limitations } \label{sec:limitation}

The study about student synchrony helps to determine the physiological response of students during different sessions of the same course. For example, if there is a significant difference in the physiological synchrony of the same student for the same course at different timings, it would be because of external factors such as the time of the day and the nature of the course. It could also suggest the difference in the teaching methodologies and the content of the course. Similarly, differences in the physiological signal of each course are also associated with the specific nature of the courses.

Although our research shows a physiological signal difference among different courses, it is important to acknowledge certain limitations. Firstly, our analysis only considers a limited number of signals for some courses. In order to better understand the connection between physiological responses and student engagement it is necessary to employ an equal number of physiological signals across all the courses. When analyzing the null hypothesis  “There is no statistically significant difference in the level of physiological synchrony among students across different types of academic courses", we found only a limited number of significant differences. Further investigation is necessary to fully comprehend the underlying factors that contribute to these variations. Finally, the study investigates physiological synchrony across different courses; however, it does not completely consider the unique characteristics of each course. Future research work could focus on specific course attributes that impact physiological responses. 

\section{Conclusion} \label{sec:conclusion}

In this study, we present our findings on physiological synchrony and physiological signal variance among various academic courses for students in distinct courses. We analyzed them using the physiological signal collected from 23 high school students for 10 different courses over the period of 4 weeks. The observed physiological synchrony score indicates a lower synchrony score for Assembly courses and a higher score for PE courses. We have observed that the degree of physiological synchrony among students varies across different courses at the intra-student level. These differences in physiological synchrony can be attributed to both individual differences among the students and the cognitive nature of the course.  In addition to the synchrony score, the courses also show distinct differences in EDA and HR features, especially Assembly and PE exhibit varied patterns. Moreover, we found that the activity of the PNS and the SNS varies depending on the nature of the course. The study suggests that students may respond differently to the demands and characteristics of specific courses, resulting in distinct physiological patterns. In conclusion, our research provides valuable insights into the complex relationship between physiological synchrony, individual differences, and the diverse nature of academic courses.

\section{Acknowledgments}

This work was supported by the ARC Centre of Excellence for Automated Decision-Making and Society (CE200100005) and the Cisco chair program of the National Industry Innovation Network (NIIN). Additionally, this work was supported by the Natural Science Foundation of China (Grant No. 62302252) and the China Postdoctoral Science Foundation (Grant No. 2023M731949). 


\bibliographystyle{ACM-Reference-Format}
\balance

\bibliography{sample-base}

\end{document}